\documentclass[reprint,
nofootinbib,
 amsmath,amssymb,
 aps,
]{revtex4-1}
\usepackage{hyperref}
\usepackage{float}
\usepackage{graphicx}
\usepackage{dcolumn}
\usepackage{bm}
\usepackage{color}      

\newcommand{\f}{\frac}
\newcommand{\h}{\hspace{0.5mm}}

\newcommand{\ds}{\displaystyle}

\begin{document}

\preprint{$\begin{array}{c}\text{UTTG-15-14}\\\text{TCC-015-14}\end{array}$}

\title{Supersymmetric Partially Interacting Dark Matter}

\author{Willy Fischler}
\email{fischler@physics.utexas.edu}
\affiliation{Department of Physics and Texas Cosmology Center\\ The University of Texas at Austin,
TX 78712.}
\author{Dustin Lorshbough}
\email{lorsh@utexas.edu}
\affiliation{Department of Physics and Texas Cosmology Center\\ The University of Texas at Austin,
TX 78712.}
\author{Walter Tangarife}
\email{wtang@physics.utexas.edu}
\affiliation{Department of Physics and Texas Cosmology Center\\ The University of Texas at Austin,
TX 78712.}

\begin{abstract}
We present a model of partially interacting dark matter (PIDM) within the framework of supersymmetry with gauge mediated symmetry breaking.  Dark sector atoms are produced through Affleck-Dine baryogenesis in the dark sector while avoiding the production of Q-ball relics.  We discuss the astrophysical constraints relevant for this model and the possibility of dark galactic disk formation. In addition, jet emission from rotating black holes is discussed in the context of this class of models.
\end{abstract}


\maketitle







\section{Introduction}

Measurements within the standard model of cosmology indicate that $27\%$ of the energy in our universe today corresponds to dark matter (DM) \cite{Ade:2013ktc,Ade:2013zuv}. In the context of $\Lambda$CDM, this matter is assumed to be cold, stable, and weakly interacting with ``visible'' matter.  Current experimental evidence for such a form of matter relies exclusively on its gravitational properties.

Models of interacting dark matter are favored by current galactic halo mass distribution models \cite{Spergel:1999mh, Dave:2000ar}. Indeed, simulations containing only collisionless dark matter show that the halo distribution is a Navarro-Frenk-White (NFW) profile with a cusp at the center. It is, then, a challenge for theorists to describe a microscopic theory of dark matter that is both interacting and in agreement with all known constraints from astrophysics, cosmology, direct/indirect detection, and collider physics.  

We present an elementary model for the composition of dark matter in our universe within the framework of supersymmetry.  Our construction is inspired by the model studied in \cite{Fischler:2010nk}. That model contains an additional unbroken gauge $U_D(1)$. We show that this scenario can have a sub-dominant ``interacting" dark matter component and still be consistent with astrophysical and cosmological constraints. For the sake of simplicity, we consider the limiting case where the dark matter couples negligibly to the standard model. This trivially satisfies constraints coming from direct/indirect detection and collider physics. 

Models where a sub-dominant component of dark matter is interacting and may form bound states, while the dominant component is collisionless, are referred to as Partially Interacting Dark Matter (PIDM) and were first considered in \cite{Fan:2013yva,Fan:2013tia}. It has been shown that in these models it is possible to form galactic disks composed entirely of sub-dominant dark matter particles \cite{Fan:2013yva,Fan:2013tia}. These models also account for the decrease of black hole angular momentum through the emission of dark sector jets \cite{Banks:2014rsa}, as well as other phenomenological consequences \cite{McCullough:2013jma,Fan:2013bea}.

Previous studies of a dark sector $U_D(1)$ gauge group and its corresponding phenomology have been carried out \cite{Holdom:1985ag,Demir:1999vv,Feng:2008mu,Ackerman:mha,ArkaniHamed:2008qn,Feng:2009mn,Kaplan:2009de,McDermott:2010pa,Fischler:2010xz,Kaplan:2011yj,Cline:2012is,Belostky:2012zz,CyrRacine:2012fz,Fan:2013yva,Fan:2013tia,McCullough:2013jma,Fan:2013bea,Banks:2014rsa}. The purpose of this study is to present a concrete microscopic framework from which one may obtain PIDM models. We additionally discuss how astrophysical and cosmological phenomenology constrains this model and yields observables in the limiting case of a negligible coupling between the dark sector and the visible standard model.

The structure of this article is the following: In section \ref{sec:model}, we present the description of our model including the calculation of the photino thermal relic density. In section \ref{ADM}, the asymmetry generation is studied using the Affleck-Dine mechanism. In section \ref{section:astro}, we apply astrophysical constraints to our model, and, in section \ref{GRB}, we analyze some consequences of PIDM for gamma ray bursts. We end this work with concluding remarks.




\section{Hidden supersymmetric sector}\label{sec:model}

In this study, we want to explore an example of a hidden sector containing stable neutral particles and particles that are charged under some ``dark" gauge interaction. The main goal is to provide a working model where the dominant constituent of dark matter (DM) is a stable, light, neutral fermion ({\it collisionless DM}), while there is a small fraction of DM that is composed by charged stable particles ({\it PIDM}). In subsequent sections, we will connect these types of models with astrophysical consequences and some of the features presented in \cite{Fan:2013yva,Fan:2013tia,Banks:2014rsa}. In this section, we describe the specifics of our model which is motivated by a similar construction introduced in \cite{Fischler:2010nk}. We consider a set of superfields that are multiplets of a dark $SU_D(2)$ gauge group in a hidden sector that is decoupled from the visible sector (MSSM). One of the chiral superfields, $H$, is a triplet of the gauge symmetry and develops a vacuum expectation value that breaks the symmetry down to a dark $U_D(1)$. In addition, we assume that the soft masses of this sector are much smaller than the soft parameters of the MSSM. At a high energy scale, there is a set of chiral superfields (messengers) that are charged the MSSM gauge symmetry group, $SU(3)_C\times SU(2)_W\times U(1)_Y$, and there are some messengers that are multiplets of $SU_D(2)$. Supersymmetry breaking is therefore communicated to both sectors through gauge mediation \cite{Dine:1981gu,Nappi:1982hm,Dine:1982zb,AlvarezGaume:1981wy}. In addition, we allow a Yukawa coupling between the $SU_D(2)$ messengers and some of the fields in the hidden sector\footnote{Direct couplings between the messengers and the low energy degrees of freedom have been studied before in the literature, although not in the context of hidden sectors \cite{Dine:1983xx,Chacko:2001km,Komargodski:2008ax,Albaid:2012qk,Kang:2012ra,Grajek:2013ola,Byakti:2013ti,Craig:2013wga,Evans:2013kxa,Abdullah:2012tq,Kang:2012sy,Fischler:2013tva}.}.
 
The chiral field content in the hidden sector is displayed in Table \ref{HiddenFields}. Besides the $SU_D(2)$ triplet, there are four chiral doublets, $X_{i=1,2},\,Y_{i=1,2}$, which carry a global $U_{X,Y}^{\text{global}}(1)$ quantum number.  The superpotential in this hidden sector is given by 
\begin{equation}
W_{\rm hidden} \,=\, Z {\rm Tr} [\lambda_h H^2 - v_h^2]\, +\,m_X X_2 X_1\, +\,m_Y Y_2 Y_1, 
\end{equation} where, $H\equiv \tau^a\,H^a$, $\lambda_h$ is a coupling and $v_h$ is a parameter with units of mass. For simplicity, henceforth $\lambda_h$ will be taken to be of order 1.

On the other hand, the messenger sector is described by the superpotential
\begin{eqnarray}
W_{\rm Mess}\,&=&\, S \left( \lambda_A A A' + \lambda_C C^2 - F \right) + M_A A B\nonumber\\& &+ M_C C D + \kappa Y_1 D X_1 \label{Wmess},
\end{eqnarray} where $S$ is the spurion field whose $F-$term breaks SUSY spontaneously, $A,\,A'\,\text{ and }B$ are multiplets of the MSSM gauge group and $C,\, D$ are triplets of $SU_D(2)$. Since we wish to have small soft parameters in the hidden sector, compared to the MSSM soft terms, we assume $M_C \gg M_A$. The mixing term that connects the low energy degrees of freedom in the hidden sector with the messenger fields in the secluded sector will yield a CP-odd term in the low energy effective theory that otherwise would be zero in a minimal GMSB scenario.

\begin{center}
\begin{table}[h]
\begin{tabular}{ | l |c|c|c|}
    \hline
    Superfield & $SU_D(2)$ & $U^{\rm global}_X(1)$ & $U^{\rm global}_Y(1)$ \\
    \hline
    $H$ & $\mathbf{3}$ & 0 & 0 \\
   \hline
   $X_1$ & $\mathbf{2} $ & $ 1 $ & 0 \\
   \hline
   $X_2$ & $\bar{\mathbf{2}} $ & $ -1 $ & 0 \\
   \hline 
   $Y_1$ & $\bar{\mathbf{2}} $ & 0&$ 1 $  \\
   \hline
   $Y_2$ & $\mathbf{2} $ & 0 & $ -1 $  \\
   \hline 
   $Z$ &$ \mathbf{1} $ & 0 & 0 \\
   \hline
    \end{tabular}
    \caption{Hidden sector chiral superfields. }
    \label{HiddenFields}
\end{table}
\end{center}
    
The $SU_D(2)$ symmetry is broken down to a $U_D(1)$ by the Higgs mechanism, in which the scalar triple $H$ takes the expectation value
\begin{equation}
\langle \vec{H} \rangle\, = \,(0,\,0,\,v_h/\sqrt{\lambda_h}).
\end{equation} The resulting supersymmetric mass Lagrangian in terms of component fields
\begin{widetext}
\begin{eqnarray}
-\mathcal{L}\,\supset\,&& \frac{2\pi\alpha_Dv^2_h}{\lambda_h} W'^+_\mu W'^{\mu -} \,+\,(\frac{\sqrt{4\pi\alpha_D} v_h}{\sqrt{\lambda_h}} \psi^+ \lambda^- +{\rm \, h.c.\, })  \,+\, \sqrt{2 \lambda_h} v_h\,\psi_{H^3} \psi_Z \\ \label{LagMass} &&\,+\,( m_X \,\psi_{X_1} \psi_{X_2}\,+\, m_Y \,\psi_{Y_1} \psi_{Y_2} +{\rm \, h.c.\, } ) + \lambda_h v_h^2 |\phi_{H^3}|^2\,+\,  2 v_h^2 |\phi_{Z}|^2  \nonumber \\
&& \,+\, m_X^2(|\phi_{X_1}|^2+|\phi_{X_2}|^2)+ m_Y^2(|\phi_{Y_1}|^2+|\phi_{Y_2}|^2),\nonumber
\end{eqnarray}  
\end{widetext}
where $\alpha_D$ is the ``dark" fine structure constant. $\psi^\pm$ are the charged fermionic components of $H$ and $\lambda^\pm$ correspond to the charged gaugini. These charged fermions combine to form two chargino mass eigenstates $\tilde{C}^\pm$. The spectrum also contains a massless vector boson $W^3_{\mu, h} \equiv \gamma_h$.

The spontaneous breaking of SUSY generates soft masses for the particle spectrum via the gauge mediation mechanism. For the scalar fields, the contribution to their masses are
\begin{equation}
\tilde{m}_{\phi_{X,Y}}^2 \approx \frac{48\,\alpha_D^2}{253 \pi^2} \left(\frac{\lambda_C\, F}{M_C}\right)^2,\qquad \tilde{m}_{\phi_{H^3}}^2 \approx \frac{3 \alpha_D^2}{8 \pi^2} \left(\frac{\lambda_C\,F}{M_C}\right)^2, \label{softscalar}
\end{equation} at the messenger scale, whereas the gaugini $\lambda_{i}$ obtain a mass
\begin{equation}
M_{\lambda} \approx \frac{\alpha_D}{4\pi} \left(\frac{\lambda_C\, F}{M_C}\right).
\end{equation}

The lightest supersymmetric particle of this sector is the ``dark photino", $\chi\equiv \lambda_3$, the superpartner of the dark photon, and has a mass $m_\chi\,\equiv\,M_\lambda$, which is parametrically much smaller than the soft masses in the visible sector. This provides a very light stable particle that might be of interest as a candidate for dark matter. The Lagrangian contains the photino interactions with the $X,\,Y$ fields $\mathcal{L}\,\supset\,\cdots -i\,\sqrt{2\pi\alpha_D}\phi_{X_i,Y_i}\,\psi_{ X_i,Y_i}^\dagger \,\chi\,+{\rm \,h.c.}$. 

\begin{center}
\begin{figure}[H]
  \centering
    \includegraphics[scale=0.3]{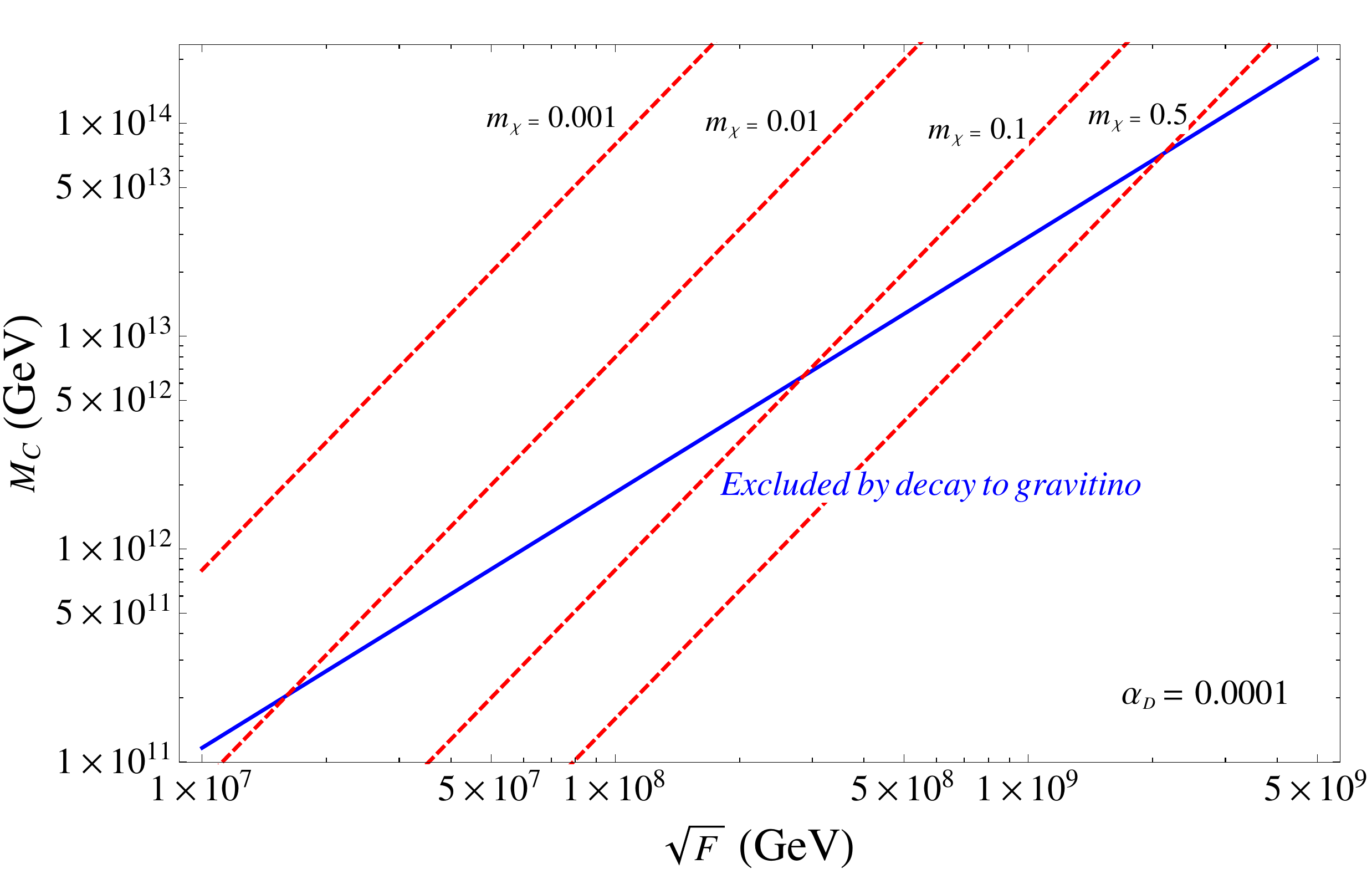}
  \caption{Different values for the photino mass in GeV as a function of $\sqrt{F}$ and $M_C$. The area under the blue line shows the region that is disallowed by requiring $\tau_{\chi} > 10^{18}$ s. }
  \label{Fig:MCvsmu}
\end{figure}
\end{center}

The dark photino can decay to a gravitino and a dark photon through the coupling of the gravitino to the supercurrent \cite{Fayet:1986zc}. Since we want the photino to be stable, we constrain its mass so that its lifetime is longer than the age of the universe. Figure \ref{Fig:MCvsmu} shows different photino masses in the $\sqrt{F} -M_C$ plane and it also depicts the region that are is discarded by requiring the photino to be cosmologically stable.

\begin{center}
\begin{figure}[H]
  \centering
    \includegraphics[scale=0.5]{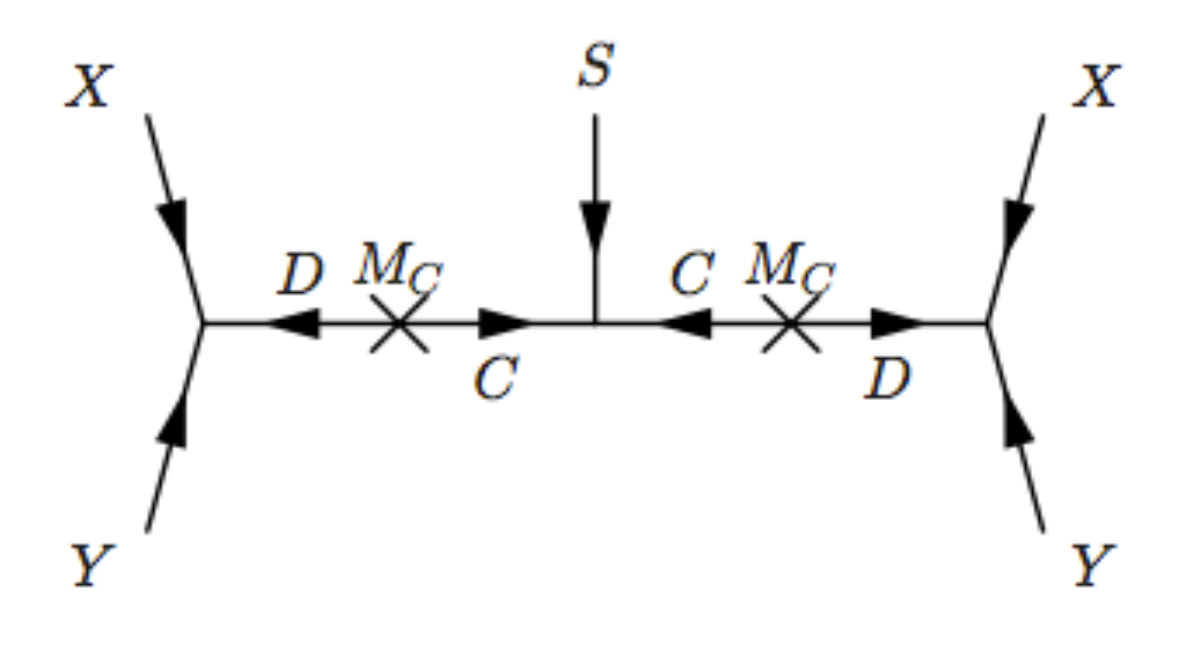}
  \caption{Feynman diagram used to generate the effective coupling $\frac{A}{M}$ in Equation (\ref{A/M}).}
  \label{Fig:FeynDiag}
\end{figure}
\end{center}

\begin{center}
\begin{figure}[H]
  \centering
    \includegraphics[scale=0.3]{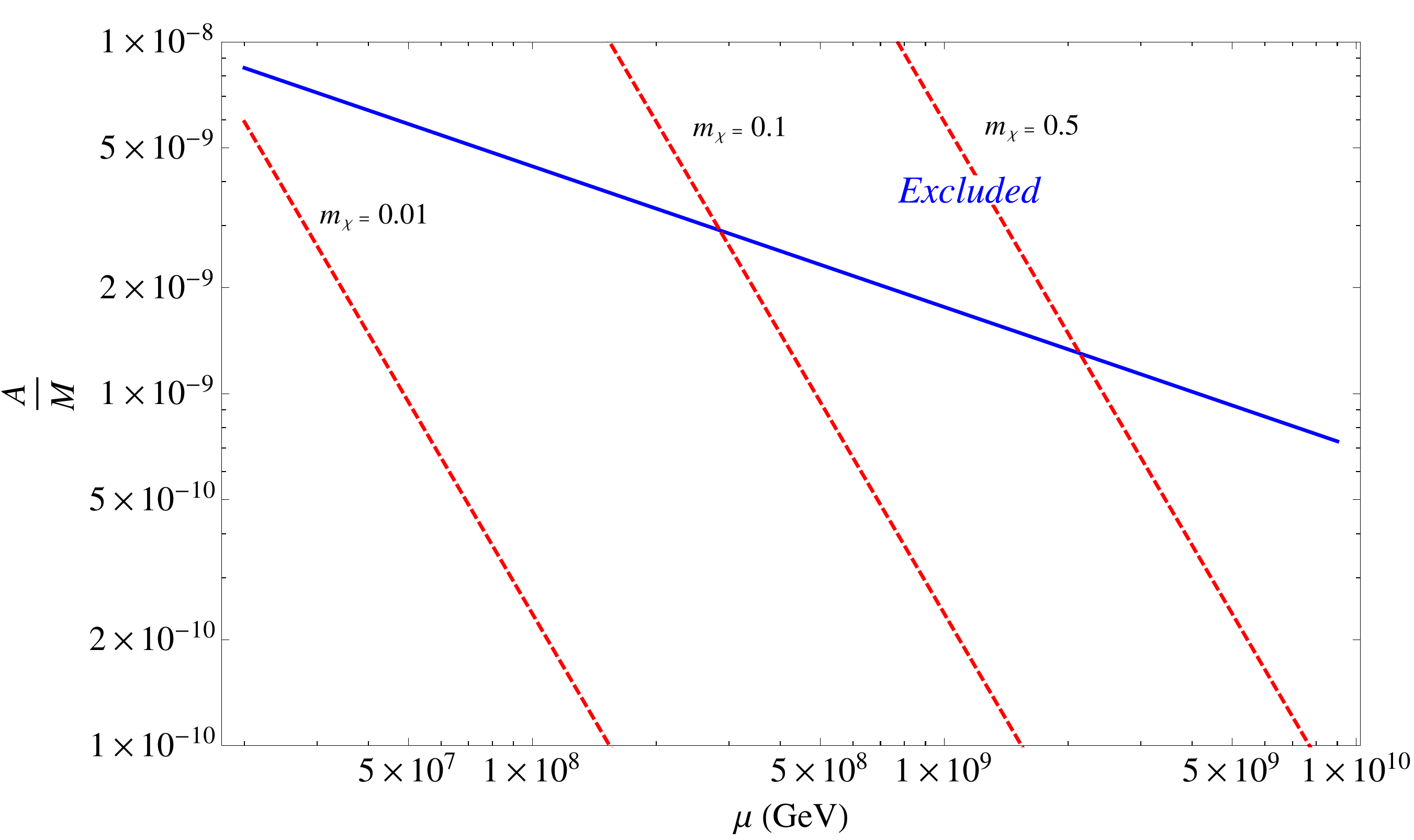}
  \caption{Values for the effective coupling $\frac{A}{M}$ as function of $\mu$ and for different values of $m_\chi$ given in GeV.}
  \label{Fig:Avsmu}
\end{figure}
\end{center}

When SUSY is broken and the messenger superfields are integrated out, the diagram shown in Figure \ref{Fig:FeynDiag} generates the effective term in the scalar potential 
\begin{equation}
V \supset \frac{A}{M} (\phi_{{X}_1} \phi_{{Y}_1})^2 \, +\, {\rm h.c.}, \label{Veff}
\end{equation}  where 
\begin{equation}
\frac{A}{M} \approx \frac{3\kappa^2 \lambda_C }{2}\frac{F}{M_C^2},\,\,\, M \approx M_C, \label{A/M}
\end{equation} which is CP-odd. This term will be used in Section \ref{ADM} to generate an asymmetry in the number densities $n_X,\,n_Y$, via the Affleck-Dine mechanism \cite{Affleck:1984fy}, since it breaks the global symmetries $U_{X,Y}^{\text{global}}(1)$. Figure \ref{Fig:Avsmu} shows different values of $\frac{A}{M}$ in terms of $\mu\equiv \sqrt{F}$ and for different values of $m_\chi$; we also have used $\kappa =1$ for simplicity. The region above the blue line corresponds to the values of $\mu$ and $m_\chi$ that are discarded by the stability of the photino.

\subsection{Symmetric dark relics}

In this section, we now study the relic density of the lightest supersymmetric particle of the hidden sector described above, the dark photino. At early times, the particles are very close to thermal equilibrium, with a temperature $T_D$. We assume that the hidden and the visible sectors are decoupled and, thus, they have different temperatures. It is standard to define the ratio between the temperatures,
\begin{equation}
\xi(t)\equiv \frac{T_D(t)}{T(t)}. 
\end{equation}
This ratio is constrained by BBN physics since the dark photons and, possibly, the lightest massive particles in the hidden sector contribute to the number of relativistic degrees of freedom of the universe. The constraint on the effective number of light relativistic species, $\Delta N_\nu < 1.0$, yields \cite{Ade:2013zuv}\footnote{A less constrained experimental bound previous to the Planck collaboration findings is \cite{Feng:2008mu}
\begin{equation}
g_{D,*}(t_{BBN})\,\xi(t_{BBN})^4\,\leq 2.52 \,\,(95\% {\rm \,C.L.}).
\end{equation}
}
\begin{equation}
g_{D,*}(t_{BBN})\,\left(\frac{T_D(t_{BBN})}{T(t_{BBN})}\right)^4\,\leq 1.75 \,\,(95\% {\rm \,C.L.}).
\end{equation} 
In our case, this bound will be obeyed as long as $\xi(t_{BBN})< 0.75$. 

The number density of relic visible photons and relic dark sector photons are given by
\begin{equation}
n_{\text{v,}\gamma}\approx\f{1}{4}T^3\approx0.33\rho_{\text{v,}\gamma}^{3/4}\approx0.07\left(\f{H}{\sqrt{G_N}}\right)^{3/2},
\end{equation}
\begin{equation}\label{eq:n_D}
n_{\text{D,}\gamma}=\f{1}{4}T_D^3=\xi^3n_{\text{v,}\gamma}=0.07\xi^3\left(\f{H}{\sqrt{G_N}}\right)^{3/2}.
\end{equation}
For approximation in section \ref{section:astro}, we will assume $\xi\approx0.5$ and $\xi\approx10^{-3}$.  However, as discussed previously, this value may be as large as 0.75.

As the temperature drops below the mass of the photino, the annihilation and production rates decrease. As a consequence, the interactions of the ``dark particles" freeze out of equilibrium.  The evolution of the particle number density of $\chi$ is described by the Boltzmann equation \cite{Kolb:1990vq}
\begin{equation}
\frac{dn_\chi}{dt}+3H\,n_\chi\,=\,-\langle \sigma_{\chi \chi} v_{\small DM} \rangle \left(n_\chi^2-n_{\chi\,eq}^2\right), \label{BoltEq1} 
\end{equation}
where $H$ is the Hubble rate and $v$ is the relative velocity of the annihilating particles. $\langle \sigma_{\chi\bar{\chi}} v \rangle$ is the thermally average annihilation cross section. 

Numerical solutions to Equation (\ref{BoltEq1}) show that the``dark" temperature at which the dark photinos depart from equilibrium is given by
\begin{eqnarray}
x_{FO}&&\equiv \frac{m_\chi}{T_{D,FO}}\nonumber\\
&&\approx \xi(t_{FO})\,{\rm ln} \left(0.015 \frac{m_\chi\,G_N^{-1/2}\langle\sigma_{\chi\chi} v \rangle \xi^{3/2} }{(g_{tot,*})^{1/2} x_{FO}^{1/2}}\right), \label{Eq:xfo}
\end{eqnarray} where 
\begin{equation} 
g_{tot,*}(T) \equiv g_{*}(T)+g_{D,*}(T_D)\left(\frac{T_D}{T}\right)^4
\end{equation} is the effective number of relativistic degrees of freedom.  The relic abundance at the present is found to be 
\begin{equation}
\Omega_{\rm \chi} h^2\equiv \frac{\rho_\chi}{\rho_c} h^2 \approx \frac{1.07 \times 10^9 \,x_{FO}\,{\rm GeV}^{-1}}{g_{tot,*}^{1/2} M_{\rm Pl}(a+3b/x_{FO})},
\end{equation} with $\langle \sigma_{\chi\bar{\chi}} v \rangle \approx a +b\,v^2$.  Where $h$ is the Hubble parameter in units of $100$ ${\rm km\, s}^{-1}{\rm Mpc}^{-1}$ and $\rho_c$ is the critical density.

\begin{figure}[H]\centering
\bigskip
\includegraphics[width=3in]{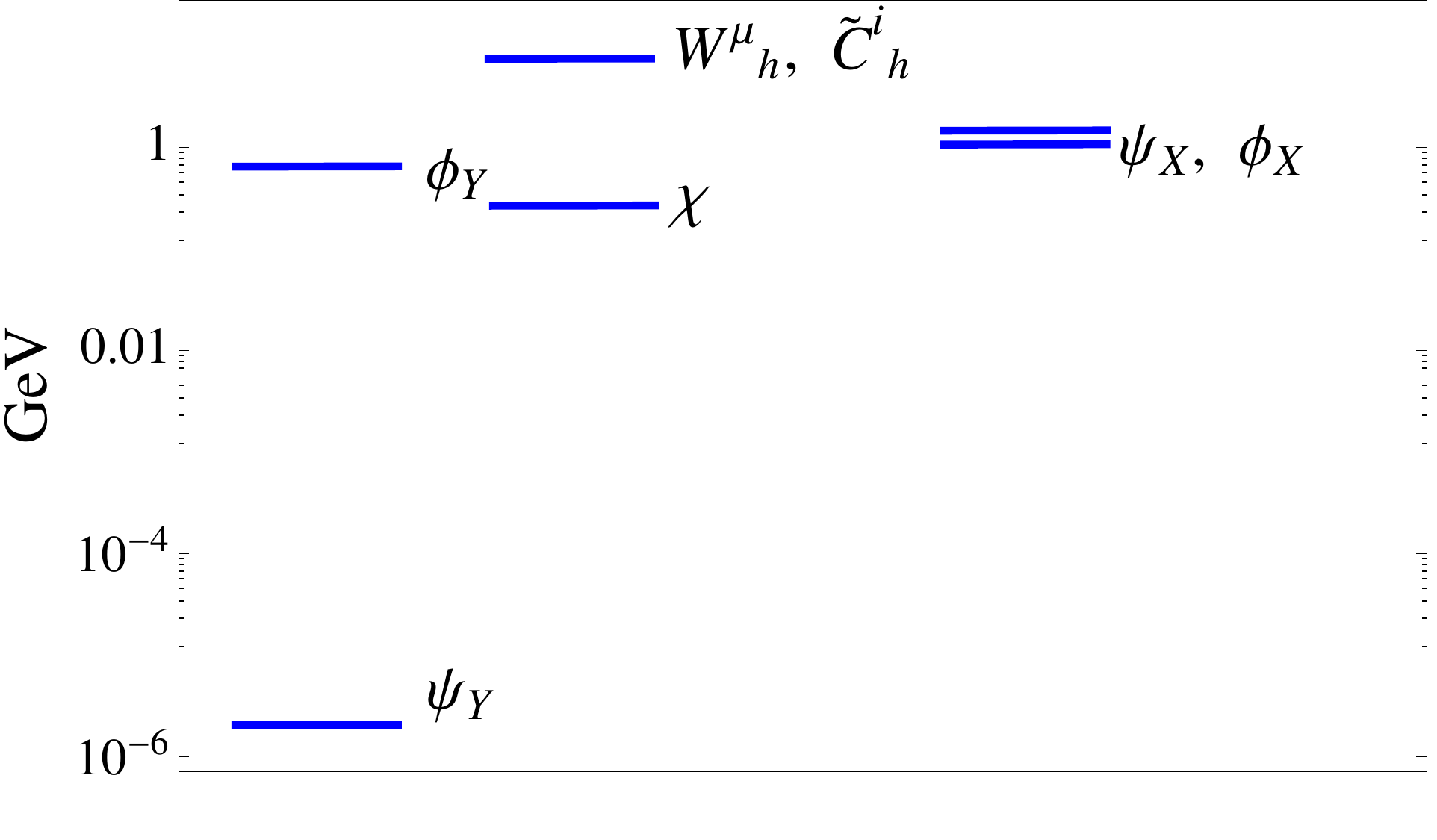}
\caption{\small Spectrum exemplifying the scenario considered in this study.} \label{fig:Spectra}
\bigskip
\end{figure}

\begin{center}
\begin{figure}[H]
  \centering
    \includegraphics[scale=0.3]{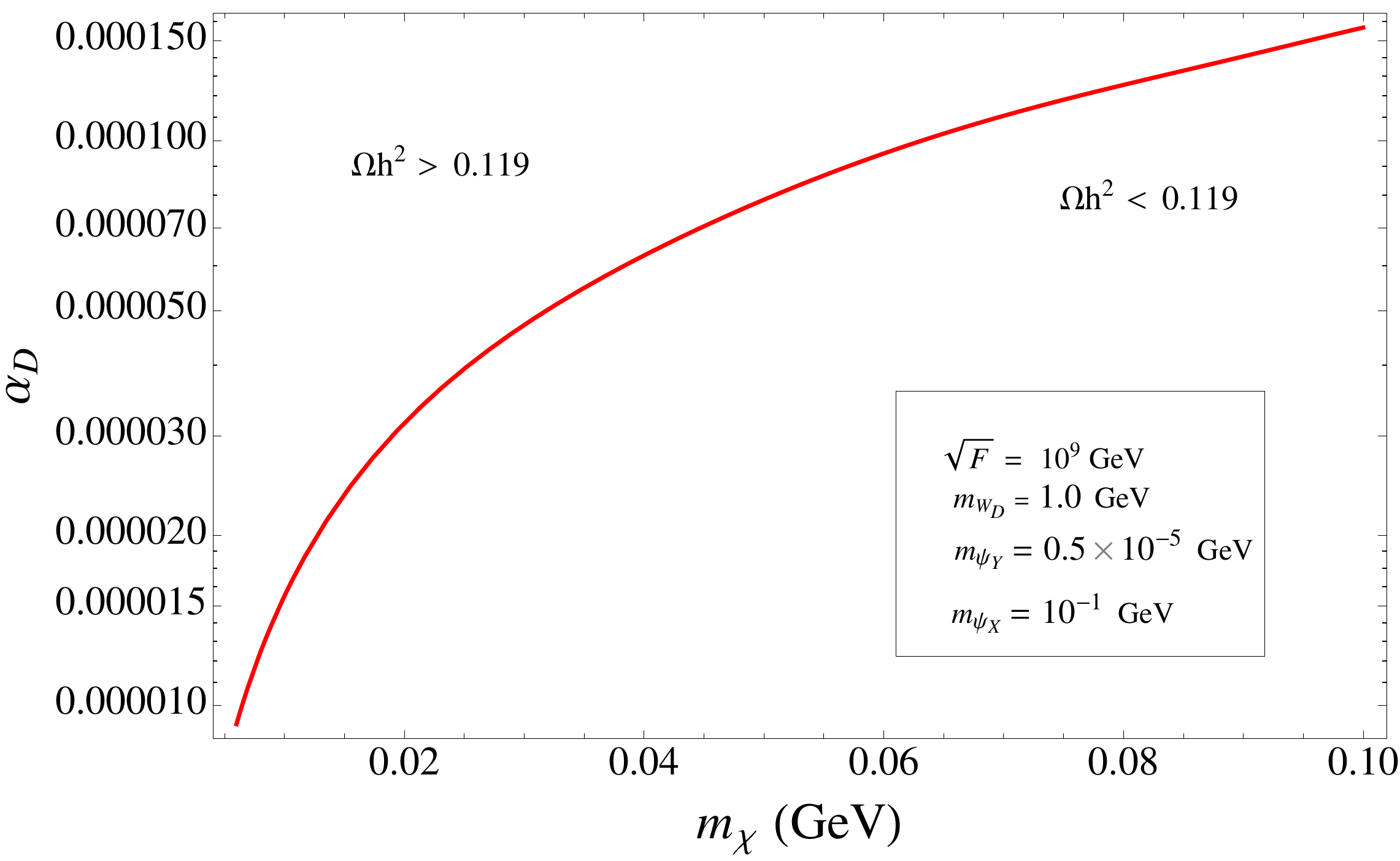}
  \caption{Range of values of $\alpha_D$ and $m_\chi$ for which $\Omega h^2\approx 0.119$ for fixed values of $m_Y$, $F$ and $v_h$ in the regime where $m_Y\ll m_\chi\ll m_X$. }
  \label{Fig:gvsmchi_I}
\end{figure}
\end{center}

In this work, we want to focus on the region of parameter space where $m_{\psi_X} \geq m_\chi > m_{\psi_Y}$. In fact, we take $\psi_i$ to have a mass under 1 MeV. As it will be shown in section \ref{section:astro}, this regime is interesting since it leads to ``dark" structure formation when there is an asymmetry in the $X,\,Y$ fields and this charged matter constitutes a small portion of the dark matter content of the universe. The process that is most relevant for the annihilation of the photini is $\chi \bar{\chi} \rightarrow \psi_{Y_i}\bar{\psi}_{Y_i}$. The  thermal averaged cross section is given by $\langle \sigma_{\chi\bar{\chi}} v \rangle \,\approx \,a +b\,v^2$, where 
\begin{widetext}
\begin{eqnarray}
a &=& \frac{3 \pi\alpha_D^2 m_Y^2 \sqrt{1-\frac{m_Y^2}{m_\chi^2}}}{8  \left(\tilde{m}_{\phi_{Y}}^2 +m_\chi^2 \left(1-\frac{m_Y^2}{m_\chi^2} \right)\right)^2}, \label{abexpansion} \\
b &=& \frac{\pi\alpha_D^2 m_\chi^2 }{64  \sqrt{1-\frac{m_Y^2}{m_\chi^2}} \left(\tilde{m}_{\phi_{Y}}^2+m_\chi^2 \left(1-\frac{m_Y^2}{m_\chi^2}\right)\right)^4}\times \left[\tilde{m}_{\phi_{Y}}^4 \left(\frac{13 m_Y^4}{m_\chi^4}-\frac{26 m_Y^2}{m_\chi^2}+16\right) \right.  \nonumber \\&& \left. -2 \tilde{m}_{\phi_{Y}}^2 m_Y^2 \left(\frac{13 m_Y^4}{m_\chi^4}-\frac{35 m_Y^2}{m_\chi^2}+22\right)+m_\chi^4 \left(\frac{m_Y^2}{m_\chi^2}-1\right)^2 \left(\frac{13 m_Y^4}{m_\chi^4}-\frac{10 m_Y^2}{m_\chi^2}+16\right)\right], \nonumber
\end{eqnarray} 
\end{widetext}
which, in the limit where $m_Y \ll m_\chi$ depends mainly on the values of the coupling constant $g_D$ and the photino mass $m_\chi$. It is noteworthy that, since the photino particles form a Majorana fermion, the $s-$wave contribution to the thermally averaged cross section is suppressed respect to the $p-$wave contribution \cite{Fayet:2004bw}.  The numerical solution to Equation (\ref{Eq:xfo}) for $m_\chi\approx 0.1$ GeV and $\xi(t_{FO}) \approx 0.6$  results in $x_{FO}\approx 10$ which is consistent with the known fact that thermal relics freeze out earlier in hidden sectors with a temperature than $T_{\rm visible}$ \cite{Feng:2008mu}. The parameter values that yield a photino relic abundance $\Omega h^2\approx 0.119$ are shown in Figure \ref{Fig:gvsmchi_I}. For the specific ranges $10^{-5} \leq \alpha_D \leq 2 \times 10^{-4}$ and $0.01\,{\rm GeV} \leq\chi \leq 0.1\,{\rm GeV}$,  the expected DM relic abundance is obtained. A case of special interest is when $\alpha_D\approx 10^{-4}$ and $m_Y\approx 5\times10^{-6}\,{\rm GeV}$, in which case a photino with a mass  $\chi\approx 0.06\,{\rm GeV}$ provides the DM relic density and will be used when discussing the astrophysical constraints in section \ref{section:astro}. 

The $X$ particles can annihilate into a pair of dark photini or into a pair of dark photons. In the case of annihilation into photini, the cross section is of the same order as the annihilation cross section of photini into $Y$ fields (Equation \ref{abexpansion}). However, the annihilation into photons is given by a cross section 
\begin{equation}
\langle \sigma_{\psi_X\bar{\psi_X}} v_X \rangle \,\approx \frac{\alpha_D\,\pi}{2\, m_X^2}\left(1-\frac{v_X^2}{8}\right), 
\end{equation} which, for the values mentioned in the previous paragraph, turns about to be about 3 orders of magnitude larger than the photino annihilation cross section. Thus, unless there is an asymmetry of these particles, the $X$ fields annihilate and there are no thermal relics of this type.

Note that if we did not require the photino to be the dominant dark matter component and instead had a different field which is completely decoupled from $X$ and $Y$, we could manipulate the values of the couplings and masses more freely and consider the regimes of parameter space discussed in \cite{Fan:2013yva,Fan:2013tia}.  We do not expand upon this case here since the primary motivation of this study is to present a self-consistent supersymmetric embedding of the PIDM scenario with minimal additional degrees of freedom.  Allowing an additional degree of freedom, such as an axion, to be the dominant dark matter while retaining the field content studied here would allow for additional freedom and potentially new phenomenology which we do not explore here.




\section{Asymmetric charged dark matter}\label{ADM}

\subsection{Asymmetry generation}

We now present the specifics of generating a net number density of charged X and Y particles. We follow the mechanism presented by Affleck and Dine \cite{Affleck:1984fy,Dine:1995kz,Allahverdi:2012ju}, in which baryogenesis was realized by flat directions of quarks.  Leptons are given some large expectation value and their subsequent evolution generates a non-zero baryon or lepton number.  In this study, we use a similar construction in order to generate non zero $X_1$ and $Y_1$ density numbers, $n_X^{\text{global}},\,n_Y^{\text{global}}$.\footnote{Some works have used the Affleck-Dine mechanism to generate asymmetries dark matter models together with baryon densities in the visible sector \cite{Bell:2011tn,Cheung:2011if,vonHarling:2012yn}.  A transfer of asymmetry from the dark sector to the visible sector may proceed even when the dark sector baryogenesis results from a first order phase transition \cite{Shelton:2010ta}.}

In Equation (\ref{Veff}), an effective term was presented after integrating-out the heavy messenger field $C$.  The potential for the $X_1$ and $Y_1$ scalar fields becomes 
\begin{equation}
V_{X,Y} = M_{X}^2 |\phi_{X_1}|^2 +M_{\tilde{Y}}^2 |\phi_{Y_1}|^2 + \left(\frac{A}{M} (\phi_{X_1}\,\phi_{Y_1})^2 \, +\, {\rm h.c.}\right), \label{Vxy}
\end{equation} where $M_{X,Y}^2$ includes the contributions to the mass of $\phi_{X_1,Y_1}$ coming from the supersymmetric Lagrangian (\ref{LagMass}) and from SUSY breaking terms (\ref{softscalar}),
\begin{equation} 
M_{X,Y}^2=m_{X,Y}^2+\tilde{m}_{\phi_{X,Y}}^2.
\end{equation} For the specific value $m_X \approx 0.1$ GeV and $m_Y \ll M_\lambda < m_X$, $M_X \sim m_X$ and $M_Y\sim \tilde{m}_{\phi_{Y}}$. 

Now, let us assume that at early times, the scalars $\phi_{X_1}$ and $\phi_{Y_1}$ had non-zero expectation values parametrized by the complex scalar $\varphi$, 
\begin{equation}
\langle \phi_{X_1} \rangle \,=\, \left( \begin{array}{c} \varphi \\ 0 \end{array} \right),\,\qquad \langle \phi_{Y_1} \rangle \,=\, \left( \begin{array}{c} 0 \\ \varphi \end{array} \right),
\end{equation}  which is a $D-$flat direction of the system. Thus, the potential in Equation (\ref{Vxy}) becomes 
\begin{equation}
V\,\equiv \, m_{\varphi}^2 |\varphi|^2 + \left(\frac{A}{M} \varphi^4 \, +\, {\rm h.c.}\right), \label{Vflat}
\end{equation} where we have redefined $m_\varphi^2 \equiv M^2_{X}\,+ M^2_{Y}$.  For the parameters considered in this paper, we will assume $m_Y\ll m_X$ and therefore $m_\varphi\approx m_X$.

The evolution equation for the scalar field in an FRW spacetime may be solved numerically for the case of radiation domination, $H=(1/2t)$.  In particular, we solve the system of equations
\begin{equation}
\ddot{\varphi}+\f{3}{2t}\h \dot{\varphi}+V_{,{\varphi^*}}=0\text{ and }\ddot{\varphi}^*+\f{3}{2t}\h \dot{\varphi}^*+V_{,{\varphi}}=0.
\end{equation}
We have chosen the initial conditions such that the field begins rolling at time $t_*^{-1}=2H_*\sim4\h\sqrt{\f{A}{M}}\h|\varphi_0|$.  This implies the initial conditions $\dot{\varphi}(t_*)=0$ and $\varphi(t_*)=|\varphi_0|\h e^{i\theta_0}$.

Let a $n_\varphi=\h i\h (\dot{\varphi}^*\varphi-\varphi^*\dot{\varphi})$ be the dark baryon/lepton number density.  The evolution in time may be easily solved.  For the case of radiation domination, $a(t)\sim t^{1/2}$, we obtain\\
\begin{equation}
\dot{n}_\varphi+3\h H\h n_\varphi=\f{1}{a^3}\f{d}{dt}\left(a^3n_\varphi\right)=2\h\text{Im}\left[V_{,\varphi}\h\varphi\right],
\end{equation}
\begin{equation}\label{eq:nint}
t^{3/2}\h n_\varphi(t) = 2\h\int_{t_*}^t\h dt'\h t'^{3/2}\h\text{Im}\left[V_{,\varphi}(t')\h \varphi(t')\right].
\end{equation}
The only term contributing to the imaginary part of (\ref{eq:nint}) is the quartic piece, 
\begin{equation}\label{eq:Im}
\text{Im}\left[V_{,\varphi}(t)\h \varphi(t)\right]=4\f{A}{M}\text{Im}\left(\varphi^{4}\right)
\end{equation}
To finish computing the integral we must choose values for the specific parameters.  For the case of (A/M) we choose $8.5\times10^{-11}$ based on the discussion in section \ref{sec:model}.  For the mass parameter we will take $m_\varphi=0.1\text{ GeV}$.  The initial phase we will choose to be such that $e^{i\theta_0}=1$ and the initial amplitude we will choose to be $|\varphi_0|=10^{7.75}$ GeV.\footnote{For instance, if there is a finite temperature potential correction of the form $\sim T^2\varphi^2$ that is relevant at early times, the initial field value may be obtained by allowing for a reheat temperature of $\sim10^{12}\text{ GeV}$ if the field value at the end of reheating is $\sim7.7\times10^{8}\text{ GeV}$.}  This initial amplitude results in production of interacting dark matter comparable to what is allowed by astrophysical bounds as we will see later.

With this choice of parameters we integrate (\ref{eq:nint}) numerically until a time $t_{\text{final}}=1700 t_*$ to ensure no further production contributes to the result and are therefore insensitive to $t_{\text{final}}$.  Choosing $t_{\text{final}}$ to be a larger value results in the integration (\ref{eq:nint}) becoming numerically unstable.  The number of dark photons at the final time is given by (\ref{eq:n_D}) with $\xi=0.5$ and $H=1/2t_{\text{final}}$.  This yields a dark sector baryon to photon ratio of
\begin{equation}
\left(\f{n_\varphi}{n_{\text{D,}\gamma}}\right)\approx6\times 10^{-12}.
\end{equation}
The standard model baryon to photon ratio is given by $\sim10^{-9}$ \cite{PDG}, though we emphasize that the dark sector is allowed to have a baryon to photon ratio very different from that in the visible sector.  In terms of number density today we find
\begin{equation}\label{eq:ADM_bd0}
n_\varphi\approx 6\times 10^{-12}n_{\text{D,}\gamma}(H=H_{\text{today}})\approx 5\times 10^{-7}\text{ cm}^{-3}
\end{equation}
As in \cite{Fan:2013yva,Fan:2013tia}, we introduce a parameter $\epsilon$ that measures the fraction of PIDM to all dark matter globally which will also be assumed to be the ratio within galaxies. We may bound this parameter using the measured density of dark matter today and (\ref{eq:ADM_bd0}). Namely,
\begin{equation}\label{eq:ADM_bd}
\epsilon=\f{\Omega_\text{PIDM}}{\Omega_{\text{DM}}}\approx 0.04.
\end{equation}
We summarize the result of choosing a different initial amplitude $|\varphi_0|$ in Table \ref{ProducedPIDM}.

\begin{center}
\begin{table}[h]
\begin{tabular}{ | l |c|c|c|}
    \hline
     $|\varphi_0|$ \text{(GeV)}& $\left(n_\varphi/n_{\text{D},\gamma}\right)$ & $\epsilon$ & $t_{\text{final}}\text{ ($t_*$)}$ \\ \hline&&&\\
    $10^{7.5}$ & $2\times10^{-12}$ & 0.01 & 700 \\
   $10^{7.75}$ & $6\times10^{-12}$ & 0.04 & 1700 \\
   $10^{8}$ & $2\times10^{-11}$ & 0.1 & 2000 \\
   $10^{8.25}$ & $4\times10^{-11}$ & 0.2 & 2000  \\
   $10^{8.5}$ & $9\times10^{-11}$ & 0.5 & 2000  \\
   \hline
    \end{tabular}
    \caption{Amount of produced interacting dark matter for different initial conditions.  We present the dark sector baryon to photon ratio, dark matter density ratio $\epsilon$ and numerically used final integration time for each initial amplitude.  The dark sector baryon to photon ratio and density ratio $\epsilon$ are insensitive to the exact value of $t_{\text{final}}$ used, though if $t_{\text{final}}$ is chosen to be too large the integration (\ref{eq:nint}) becomes numerically unstable.}
    \label{ProducedPIDM}
\end{table}
\end{center}

In principle, one may need to be concerned with the formation of Q-balls in the context of gauge mediation symmetry breaking and the Affleck-Dine mechanism \cite{Coleman:1985ki,Kusenko:1997si,Enqvist:1997si,Laine:1998rg,Kasuya:1999wu,Kasuya:2000sc,Kasuya:2001hg,Chiba:2009zu}. The effective scalar potential for $\varphi$ in our model is given by \cite{Kusenko:1997si,Enqvist:1997si}
\begin{equation}
V_{eff}(\varphi) \approx m_\varphi^4{\rm Log}\left[1+\left(\frac{|\varphi|^2}{M_C^2}\right)\right] + \lambda |\varphi|^4,  
\end{equation} where the logarithmic term comes from integrating out the messenger $D$ in Equation (\ref{Wmess}) and $\lambda \sim 10^{-24}$. Stable Q-balls exist if $V_{eff}(\varphi)/|\varphi|^2$ has a minimum for some non-zero value of $\varphi$ \cite{Coleman:1985ki,Kusenko:1997ad}. In our case, there does not exist such a minimum.

Under $U_D(1)$, the excess $\psi_{X_1}$ and $\psi_{Y_1}$ have charges $+1$ and $-1$ respectively.  This in principle allows them to form bound state atoms analogous to Hydrogen.  We emphasize that we do not have dark quarks and that $\psi_{X_1}$ is not a composite particle, but the astrophysical phenomenology of the bound state atom is still similar to that of visible sector Hydrogen.  However, for the parameters considered in this study, the interacting dark matter never decouples from the dark photon bath.  We will discuss how to interpret astrophysical constraints in this context in the next section.

The number density we have obtained, $n_\varphi$, is the global number density of the interacting Dark matter particles.  For the remainder of this study, we will be interested in the local number densities of the $n_X$ and $n_Y$, which we will take to be equal. In the next section, we will find bounds on the local number density of interacting dark matter and compare those bounds to the bound on the global number density through the parameter $\epsilon$.




\section{Astrophysical and Cosmological Constraints}\label{section:astro}
The model we have proposed will now be shown to be consistent with bounds arising from astrophysical and cosmological considerations.  Following \cite{Fan:2013yva,Fan:2013tia}, we will consider halo shape analysis and the bullet cluster observation.  We will additionally discuss the impact of dark acoustic oscillations \cite{Cyr-Racine:2013fsa}.

There is a cosmological constraint arising from the existence of dark acoustic oscillations \cite{Cyr-Racine:2013fsa}.  Radiation pressure due to the interacting dark matter opposes matter infall and results in shallower gravitational potential wells which impact the CMB and matter power spectrum.  For the benchmark parameters discussed in section \ref{sec:model} with $\xi=0.5$, this corresponds to a bound of $\epsilon\lesssim 0.04$\footnote{We would like to thank Francis-Yan Cyr-Racine for bringing this to our attention in a private communication.}. This constraint will decrease for the case of lower $\xi$.  This constraint is the most restrictive for both the case in which galaxies form and the case in which galaxies do not form.  We have shown that our benchmark parameters readily satisfy this bound in (\ref{eq:ADM_bd}).

There are two regimes of parameter space given the benchmark parameters discussed in \ref{sec:model}.  One for which the PIDM does not decouple and remains a dark plasma today, and one for which the PIDM does decouple and dark atoms may be long lived and form galaxies.  For the case of $\xi\approx0.5$, the dark photon temperature is always higher than the binding energy of the dark atoms and therefore the PIDM never decouples from the dark photon bath.  For the case of a much colder dark sector than the visible sector, $\xi\lesssim 10^{-2}$, the binding energy is eventually larger than the dark photon bath temperature and we may produce long lived dark atoms.  However, the parameters which allows for dark galaxy formation results in a large $\Sigma_{DAO}$ ($\sim10-100$) for the dark acoustic oscillation analysis presented in \cite{Cyr-Racine:2013fsa} and is likely severely constrained.  For the remainder of our study we focus on the case in which dark decoupling does not occur.

\subsection{The Bullet Cluster and Halo Shape Analysis}
There are potentially two constraining astrophysical bounds for the case in which dark galaxies do not form, the Bullet Cluster and halo shape analysis.  The constraint on the amount of PIDM, $\epsilon$, comes from observations of Bullet Cluster \cite{Clowe:2003tk,Markevitch:2003at}.  In particular, from measurements of the mass-to-light ratio of the cluster and subcluster one can obtain an upper bound on the fraction of dark matter lost in the galactic merger.  The particle loss fraction is determined by the fractional decrease of the mass to light ratio (M/L) for the subcluster within 150 kpc,
\begin{equation}
f=\f{|\left(M/L\right)_{\text{I,main cluster}}-\left(M/L\right)_{\text{I,subcluster}}|}{\left(M/L\right)_{\text{I,main cluster}}}.
\end{equation}
Here the subscript I denotes the Ith frequency band chosen for determining (M/L).  Lensing map analysis \cite{Bradac:2006er,Randall:2007ph} has determined $\left(M/L\right)_{\text{I,subcluster}}=179\pm11$ and $\left(M/L\right)_{\text{I,cluster}}=214\pm13$ which results in an upper bound of the particle loss fraction of $f\lesssim0.30$ \cite{Randall:2007ph} to 95\% confidence.

In the scenario we are considering where the dominant component of dark matter is collisionless, the particle loss fraction bound becomes a bound on the amount of dark matter that can be interacting, $\epsilon\lesssim0.30$.

The second constraint from astrophysics confronting our model is that of recent NGC 720 halo ellipticity measurements.  By modeling the galaxy as a pseudoisothermal distribution, the deviation from sphericity was found to be $35\%$ at 5-10 kpc from the galactic center \cite{Buote:2002wd}. Constraints on the self interaction cross section of dark matter due to ellipticity have been studied both analytically \cite{MiraldaEscude:2000qt} and numerically \cite{Peter:2012jh}.  The self interaction cross section constraints are not relevant for the PIDM we discuss in this study since we require the dominant component to be collisionless, rather what one may ascertain is the allowed number of dark matter interactions allowed in the lifetime of the universe.  As discussed in \cite{Fan:2013yva,Fan:2013tia}, PIDM scenarios seem to readily satisfy the constraints on total number of dark matter interactions within the lifetime of the universe as shown in figure 5 of \cite{Peter:2012jh}.

For the specific case of our benchmark parameters from section \ref{sec:model} and $\xi=0.5$, the strongest constraint on $\epsilon$ is therefore the constraint arising from dark acoustic oscillation considerations previously discussed in which $\epsilon\lesssim 0.04$.

\section{Gamma Ray Burst Phenomenology}\label{GRB}
In a companion paper \cite{Banks:2014rsa}, we have explored the possibility of dark matter which is neutral under the standard model modifying our conclusions about black hole spin measurements during gamma ray burst emission events.  The main result from that study is that the rate of change of the dimensionless spin parameter $\ds{-1<a\equiv J/G_N\h M_B^2<+1}$, for the case of prograde rotation ($a>0$), may be written in terms of the black hole mass $M_B$, visible infall $\dot{M}_{\text{in,v}}$, the visible jet emission $L_{\text{jet,v}}$, the dark matter infall $\dot{M}_{\text{in,D}}$, the dark matter jet emission $L_{\text{jet,D}}$, and the gravitational radiation emission $L_{\text{gr}}$ as
\begin{equation}\label{eq:a_dot}
\dot{a}=\f{\lambda\h\gamma}{M_B}\h\left(\dot{M}_{\text{in,v}}+\dot{M}_{\text{in,D}}-L_{\text{jet,v}}-L_{\text{jet,D}}-L_{\text{gr}}\right).
\end{equation}
We have defined $\ds{\lambda\equiv2\sqrt{2}\left(\f{1-a^2}{1-\sqrt{1-a^2}}\right)^{1/2}}$ and $\ds{\gamma\equiv\f{1}{\sqrt{2}}\left(1+\sqrt{1-a^2}\right)^{1/2}}$.  In \cite{Banks:2014rsa} we have given numerical estimates for these terms, but ultimately ``two-sector" numerical simulations should be done to develop a better understanding of how such a sector may modify the expected change in spin during any such event.  We have assumed the increase of irreducible black hole mass during infall is negligible.  In order to justify this assumption, note that the change in irreducible mass is proportional to the black hole temperature, $T_{BH}$, which vanishes for spin parameter near unity
\begin{eqnarray}
\delta M_{\text{irr}}&&=\f{1}{4}\h M_P^2\h T_{BH}\h\delta A_{\text{horizon}}\nonumber\\
&&=\f{1}{16\h\pi}\f{M_P^4}{M_B}\f{\sqrt{1-a^2}}{\left(1+\sqrt{1-a^2}\right)}\delta A_{\text{horizon}}.
\end{eqnarray}
Previous numerical studies have shown that the black hole spin rapidly grows during the collapsing stage \cite{MacFadyen:1998vz}, and therefore the irreducible mass becomes approximately constant after a short time.

The scenario described in \cite{Banks:2014rsa} relies upon the assumption that the collapsar model \cite{Woosley:1993wj,Paczynski:1997yg,MacFadyen:1998vz} is sufficient to explain some of the observed long gamma ray bursts and that the jets themselves are dominantly generated by the Blandford-Znajek (BZ) mechanism \cite{Blandford:1977ds,Komissarov01,Komissarov:2002dj,Komissarov:2004qu,Koide2004,McKinney:2004ka,McKinney:2006tf,Barkov:2007us,Komissarov:2008yh,Barkov:2008wc,Nagataki:2009ni,Komissarov:2009dn,Tchekhovskoy:2011zx, Penna:2013rga}.  The proposal in \cite{Banks:2014rsa} is that by measuring the spin of the black hole over time and comparing with observed jet emission, simulated visible matter infall from the progenitor star, and simulated gravitational radiation losses, one may bound the amount of dark matter infall and dark jet emission.

In order for there to be dark matter in the immediate vicinity of the progenitor star and therefore non-negligible dark matter infall, we must have some localized overdensity of interacting dark matter. The dark matter gas may be dense enough to cause gravitational microlensing in which case constraints from MACHO searches become relevant.  Dense, compact objects generically are referred to as MACHOS (Massive Compact Halo Objects).  The existing constraints on MACHOS \cite{Lacey85,Yoo:2003fr,Moniez:2010zt,Iocco:2011jz} are from gravitational microlensing experiments and stability of wide binary star systems.  Gravitational microlensing occurs due to a massive compact object moving within the line of sight of an observer and a light source.  The result is that the light source is temporarily magnified.  Since present studies depend heavily on the model of the dark matter distribution, they are not directly applicable to the case of dark matter disk galaxies without further analysis \cite{Fan:2013yva}.  We note dark matter capture by the progenitor star will be small since we have not allowed for non-gravitational interactions between the dense progenitor core and the interacting dark matter.  Previous studies have shown that for these types of models, even if a small interaction between nucleons and the interacting dark matter is allowed, the amount of capture is small \cite{Fan:2013bea}.

We emphasize that it is not unreasonable to assume that there are local over-densities of interacting dark matter in the immediate vicinity of a progenitor star.  In the dwarf galaxy Mrk 996 there is evidence of a recent minor merger that has allowed for an abundance of Wolf-Rayet star formation \cite{Jaiswal:2013cya}.  In addition to acquiring baryonic matter during the merger, interacting dark matter would be acquired as well.  Dwarf galaxies tend to be overwhelmingly dominated by dark matter, therefore we naively expect there to be approximately four times as much interacting dark matter in a dwarf galaxy than baryonic matter ($\epsilon\lesssim4\%$).

Dynamically, as the merger occurs the baryonic matter and the interacting dark matter will follow that same orbital path since they interact identically gravitationally.  As the baryonic matter clumps, some of the interacting dark will remain stuck in the resulting gravitational wells.  Concretely, we treat the interacting dark matter plasma as a collapsing cloud of self-gravitating gas around the baryon induced gravitational well.  In order for the density of the gas to be comparable to the density of the baryonic accretion disk in the collapsar model ($\sim 10^{6\pm2}\text{ g/cm}^3$ \cite{MacFadyen:1998vz}) the Jeans mass of the gas must be\footnote{The Wolf-Rayet stars in Mrk 996 are $\sim4.5$ Myr old ($z\sim10^{-4}$) \cite{James2009}, therefore the relevant interacting dark matter temperature is that of the dark photon bath approximately today since the interacting dark matter is still coupled with the dark photon bath.} 
\begin{equation}
M_J\approx\sqrt{\f{375}{4\pi}}\left(\f{k_BT_{D}(0)}{G_Nm_X}\right)^{3/2}\f{1}{\sqrt{\rho_{\text{acc}}}}\approx 1.3\times 10^{-11\pm1}\text{ M}_{\text{Sun}}.
\end{equation}

For the specific case of Mrk 996, the total amount of all dark matter is $\sim 10^8\text{ M}_{\text{Sun}}$ \cite{James2009} and therefore the total amount of allowed interacting dark matter is $\lesssim 10^{6}M_{\text{Sun}}$.  This allows for many such clouds of interacting dark matter to form even if most of the interacting dark matter sinks to the center of the galaxy.  Further analysis for other galaxies and an extension to the case of decoupled interacting dark matter will be addressed in future work.

The amount of visible matter infall may be calculated using the free-fall model \cite{Cooperstein84,Bethe:1990mw,Komissarov:2009dn}.  We review this argument here in order to comment on its applicability to the dark sector.  In the visible sector the pressure, $P_{\text{v}}$, for the gas we consider is dominated by electrons since nucleons are in large nuclei.  The equation of state for the relativistic electrons is given by
\begin{equation}
P_{\text{v}}=K_{\text{v}}\h\rho_{\text{v}}^{4/3},
\end{equation}
where $K_{\text{v}}=\f{1}{4}\epsilon_F\h Y^{4/3}_e\left(1+\f{2}{3}\left(\f{S_e}{\pi}\right)^2\right)\rho_{\text{v}}^{-1/3}.$  We have introduced the electron to nucleon ratio $Y_e$, the electron Fermi energy $\epsilon_F=\left(3\pi^2\rho_{\text{v}} Y_e\right)^{1/3}$ and the entropy per electron $S_e=\left(\pi^2\h T/\epsilon_F\right)$.  The density distribution prior to collapse is given in terms of a mass dependent ${\mathcal{O}}(1)$ coefficient, $C_1$, by
\begin{equation}\label{eq:rho_0}
\rho_{\text{v}}(\Delta t=0)\approx 10^{31}\h C_1\h \left(\f{1\text{ cm}}{r}\right)^{3}\text{ g/cm}^3.
\end{equation}
The density and matter infall rate after an elapsed time $\Delta t$ are given as
\begin{equation}
\rho_{\text{v}}\approx 10^{55}\h C_1\h\f{4}{\sqrt{G_N\h M_B\h M_{\text{Sun}}}}\left(\f{1\text{ s}}{\Delta t}\right)\left(\f{1\text{ cm}}{r}\right)^{3/2}\text{ g/cm}^{3},
\end{equation}
\begin{equation}
\dot{M}_{\text{in,v}}\approx(0.04)\h C_1\left(\f{1\text{ s}}{\Delta t}\right)\text{ M$_{\text{sun}}$/s}.
\end{equation}

The dark sector gas infall scenario may be significantly more complicated than the scenario for the visible sector that we have presented.  The initial conditions for the dark sector gas are independent of the stellar properties of the progenitor star since it only interacts gravitationally with the visible stellar matter.  The position of the dark sector gas is influenced by the position of the visible progenitor gravitational well, but the magnetohydrodynamical properties of the dark gas are not.  How exactly the dark sector cloud of gas, which may be nearly coincident with the progenitor star, infalls requires further study of dark sector substructure which we leave for future work.

The dark $U(1)_D$ sector allows for dark electromagnetism similar to electromagnetism in the visible sector.  Therefore jet production through the BZ mechanism may proceed as is well-known for the visible sector.  Since the microscopic properties of the interacting dark matter need not be identical to those of the visible sector it may be that a given collapsar event allows jet production for the visible sector but not for the dark sector.  In particular, the mass-to-charge ratio and fine structure constant must allow for the Alfven speed to exceed the local free fall speed in the ergosphere \cite{Komissarov:2009dn} and pair production to be efficient \cite{Blandford:1977ds}.

Observations of Fe K$\alpha$ spectral emission \cite{McClintock:2011zq,Brenneman:2013oba,Fabian:2014aja} have allowed astronomers to determine spin for black holes at various redshifts.  In particular, the spin has been determined for some supermassive blackholes at redshifts comparable to those at which we observe long gamma ray bursts.  Therefore it seems to us that it is in principle possible to determine the spin of the newly formed black hole in a collapsar scenario that may underlie some long gamma ray bursts.  Studies for future missions \cite{Takahashi:2008bu,Takahashi:2010fa,White:2010bx,Barcons:2012zb,Barret:2013bna,Garcia:2011ra,Feroci:2011jc} are presently underway to further develop our capability to measure black hole spin.

\section{Conclusions}
We have described a microscopic model of PIDM within the framework of supersymmetry. We have discussed the astrophysical and cosmological constraints for such a model in the limiting case that the interactions between the dark sector and visible sector are negligible. Furthermore, we have explored ways in which this class of models may be relevant for observational studies of gamma ray bursts and explored the phenomena of dark sector jets powered by the Blandford-Znajek mechanism. Our proposal to compare spin down rate with jet emission luminosity potentially provides a new tool to study the microscopic theory of dark matter. 

The model proposed here may be generalized to larger symmetry groups or to allow a stronger coupling between the dark sector and the visible sector. These are interesting directions for future work. Another interesting question is how collapsar model physics is modified by the existence of such a dark sector.

\section*{Acknowledgments}
D.L. and W.T. would like to thank Paul Shapiro and Kathryn Zurek for helpful discussions.  D.L. would like to thank Jimmy for useful discussions.  We would like to thank Francis-Yan Cyr-Racine for helpful comments on an earlier draft of this paper.  This work was supported by the National Science Foundation under Grant Number PHY-1316033.
 
\end{document}